\documentclass[prd,aps,showpacs,preprintnumbers,amssymb]{revtex4}

\usepackage{color}
\usepackage{epsf}

\def\e3p{$\eta \rightarrow 3 \pi$}

\begin{document}
\title{%
\hfill{\normalsize\vbox{%
\hbox{}
 }}\\
{Spontaneous symmetry breaking and Goldstone theorem for composite states revisited}}

\author{Amir H. Fariborz
$^{\it \bf a}$~\footnote[1]{Email:
 fariboa@sunyit.edu}}

\author{Renata Jora
$^{\it \bf b}$~\footnote[2]{Email:
 rjora@theory.nipne.ro}}

\affiliation{$^{\bf \it a}$ Department of Matemathics/Physics, SUNY Polytechnic Institute, Utica, NY 13502, USA}
\affiliation{$^{\bf \it b}$ National Institute of Physics and Nuclear Engineering PO Box MG-6, Bucharest-Magurele, Romania}

\date{\today}

\begin{abstract}
We discuss the well-known phenomenon of spontaneous symmetry breaking for  a linear sigma model for scalar and pseudoscalar mesons based on the meson composite structure and  the normalization of the quantum  states.  To test our formulation and validate our approach we give another proof of the Goldstone theorem  and derive the corresponding mass eigenstates of the theory. We briefly describe the possible wave function of a meson that leads to the adequate mass eigenstates.

\end{abstract}
\pacs{11.30.Rd, 11.30.Hv, 12.60.Rc}
\maketitle

\section{Introduction}
It is now well known that QCD is the theory of the strong interaction. At higher energies the coupling constant is small and there is an adequate perturbative description of the theory in terms of quarks and gluons. At low energies the coupling constant is large, the theory confines  and one needs to introduce new dynamical degrees of freedom, the hadrons. The low energy hadron spectrum includes among others three very light particles, the pions. In order to explain the smallness of their masses and other properties of the hadron spectra one assumes that the corresponding meson Lagrangian is endowed with a chiral flavor symmetry: $SU(2)_L\times SU(2)_R$ (we consider only two flavors). This symmetry is both spontaneously broken by the formation of the chiral condensate as well as explicitly broken in the presence of the quark masses.  The phenomenon of spontaneous chiral symmetry breaking is evident in a linear sigma model that includes both scalars and pseudoscalars and it is a particular example of spontaneous symmetry breaking of a global or gauge group which was described in a series of pioneering works \cite{Nambu}-\cite{Hagen}.

The aim of our work is to analyze spontaneous symmetry breaking from the point of view of the composite structure of the meson states. Although determining the exact substructure of the mesons in terms of the constituent quarks can be quite nontrivial, some general features may be considered and some important properties may be derived based on those. In section II we will show what one can learn about the meson states and spontaneous symmetry breaking for composite theories by considering the meson quark substructure. In essence we give another proof of the Goldstone theorem based on the normalization of the quantum meson states.  In section III we compute the mass eigenstates of the hamiltonian  after spontaneous symmetry breaking.  Section IV contains a description of a possible  meson wave function that leads to the correct mass spectrum. Section V is dedicated to the conclusions.

\section{Composite states and the goldstone theorem}

We consider a model that contains composite scalars and pseudoscalars, the most relevant example being the $SU(2)_L \times SU(2)_R$ linear sigma model for the mesons.  The corresponding meson states have the structure \cite{Schechter}:
\begin{eqnarray}
M_{ij}\propto\bar{\Psi}_i^A(\frac{1+\gamma^5}{2})\Psi^A_j,
\label{mestr45546}
\end{eqnarray}
where $i$, $j$ denote the flavor and $A$ represents the color index. The model displays spontaneous symmetry breaking down to $SU(2)_V$ as soon as the vacuum condensates $\bar{\Psi}_i^A\Psi_i^A$ forms.  In a linear sigma model the particles arrange themselves in triplets and singlets of the flavor group according to the representation:
\begin{eqnarray}
&&\pi^+\propto(-i)\bar{d}\gamma^5u
\nonumber\\
&&\pi^0\propto i\frac{1}{\sqrt{2}}[\bar{d}\gamma^5d -\bar{u}\gamma^5u]
\nonumber\\
&&\eta\propto -i\frac{1}{\sqrt{2}}[\bar{d}\gamma^5u+\bar{u}\gamma^5d]
\nonumber\\
&&a^+\propto \bar{d}u
\nonumber\\
&&a^0\propto\frac{1}{\sqrt{2}}[\bar{u}u-\bar{d}d]
\nonumber\\
&&\sigma\propto\frac{1}{\sqrt{2}}[\bar{u}u+\bar{d}d].
\label{mes55466}
\end{eqnarray}
Here $\pi^{\pm}$, $\pi^0$ and $\eta$ correspond to the pseudoscalar states whereas $a^{\pm}$, $a^0$ and $\sigma$ to the scalar ones. We are only considering a $\bar{q}q$ model to illustrate the  connection but in reality it is known that the light scalars are considerably more complex and contain large four-quark component and glue component.

If one considers a simple Lagrangian that contains only the kinetic terms and a mass term then both the meson states in  Eq. (\ref{mes55466}) and the quantities $\bar{\Psi}_i(\frac{1+\gamma^5}{2})\Psi_j$ are eigenstates of the Hamiltonian with the same mass.

Although our arguments could extend easily to the model above in what follows we shall consider a simplified version of it with a single quark, assuming this to be $u$. The model has a simple $U(1)_L \times U(1)_R$ symmetry and contains a scalar, $\Phi_0\propto\frac{1}{\sqrt{2}}\bar{u}u$ and a pseudoscalar $\Phi_1\propto-i\frac{1}{\sqrt{2}}\bar{u}\gamma^5u$. The Lagrangian of interest is given simply by:
\begin{eqnarray}
{\cal L}=\partial^{\mu}(\Phi_0-i\Phi_1)\partial_{\mu}(\Phi_0+i\Phi_1) -m^2(\Phi_0^2+\Phi_1^2)=\frac{1}{2}\partial^{\mu}M^{\dagger}\partial_{\mu}M-\frac{1}{2}m^2M^{\dagger}M.
\label{lagr45534}
\end{eqnarray}
Note that both the fields $\Phi_0$ and $\Phi_1$ or $u_R^{\dagger}u_L$ and $u_L^{\dagger}u_R$ may be considered eigenstates of the Hamiltonian. In terms of the latter fields the Lagrangian becomes:
\begin{eqnarray}
{\cal L}\propto\frac{1}{2}[\partial^{\mu}(u_L^{\dagger}u_R)\partial_{\mu}(u_R^{\dagger}u_L)-m^2(u_L^{\dagger}u_R)(u_R^{\dagger}u_L)],
\label{lagr45546}
\end{eqnarray}
since $\Phi_0+i\Phi_1\propto \frac{1}{\sqrt{2}}u_R^{\dagger}u_L$.

Assume that a vacuum condensate of $\Phi_0$ forms:
\begin{eqnarray}
\langle|\Phi_0(x)|\rangle=\langle|\Phi_0(0)|\rangle=v
\label{v45546}
\end{eqnarray}
Although we do not know exactly the exact structure of the scalars in terms of the up quark fields we can write quite safely the analogue of Eq.(\ref{v45546}) as function of the individual quark states:
\begin{eqnarray}
\int \frac{d^3p}{(2\pi)^3}\int \frac{d^3q}{(2\pi)^3}f(\vec{p},\vec{q})\sum_{s,r}\langle 0|b^s_p\bar{v}^s(p)v^r(q)b_q^{r\dagger}|0\rangle\neq 0,
\label{resttrrtt6}
\end{eqnarray}
where $b^s_p$ and $b_q^{r\dagger}$ are operators of annihilation and creation, $v$ is the Dirac spinor solution of the equation of motion and $f(\vec{p},\vec{q})$ is a function that encapsulates our lack of knowledge with regard to the scalar structure in terms of the constituents. The result will contain a delta function $ \delta(\vec{p}-\vec{q})$ and the element $\bar{v}^s(p)v^r(q)=-2m\delta_{rs}$, where $m$ is the mass of an individual quark. It can be observed that the matrix element in Eq. (\ref{resttrrtt6}) is in general different than zero if the quark has a nonzero mass.

Let us consider the wave packets associated to the fields $\Phi_0$ and $\Phi_1$:
\begin{eqnarray}
&&|\Phi_0\rangle=\int \frac{d^3\vec{k}}{(2\pi)^3}\frac{1}{\sqrt{2E_k}}\Phi_0(\vec{k})|\vec{k}\rangle
\nonumber\\
&&|\Phi_1\rangle=\int \frac{d^3\vec{k}}{(2\pi)^3}\frac{1}{\sqrt{2E_k}}\Phi_1(\vec{k})|\vec{k}\rangle,
\label{res553443}
\end{eqnarray}
where $\Phi_0(\vec{k})$ and $\Phi_1(\vec{k})$ are the Fourier transforms of the respective spatial wave function.

Since $\Phi_0$ and $\Phi_1$ are composite they can be expressed quite generically in terms of the constituent quarks as:
\begin{eqnarray}
&&|\Phi_0\rangle=\int \frac{d^3\vec{k}}{(2\pi)^3}\frac{d^3\vec{p}}{(2\pi)^3}f(\vec{p},\vec{k})\sum_{r,s}\bar{u}^s(k)v^r(p)|\vec{k},s;\vec{p},r\rangle
\nonumber\\
&&|\Phi_1\rangle=\int \frac{d^3\vec{k}}{(2\pi)^3}\frac{d^3\vec{p}}{(2\pi)^3}f(\vec{p},\vec{k})\sum_{r,s}\bar{u}^s(k)\gamma^5v^r(p)|\vec{k},s;\vec{p},r\rangle.
\label{const5546}
\end{eqnarray}
Here the function $f(\vec{k},\vec{p})$ is as before and $u$, $v$ are the standard spinor  solutions of the Dirac equation of motion. It is necessary to have the same function $f(\vec{k},\vec{p})$ for both the scalar and the pseudoscalar in order to have the correct representation in terms of the quarks:
\begin{eqnarray}
&&\Phi_0+i\Phi_1\propto u_R^{\dagger}u_L
\nonumber\\
&&\Phi_0-i\Phi_1\propto u_L^{\dagger}u_R.
\label{res3324567}
\end{eqnarray}
Note that here as opposed to the Eq. (\ref{const5546}) $u$ designates the up quark.

Having the generic expressions both for $|\Phi_0\rangle$ and $|\Phi_1\rangle$ we shall compute the normalization of states. Thus,
\begin{eqnarray}
&&\langle \Phi_0|\Phi_0\rangle=\int \frac{d^3\vec{k}}{(2\pi)^3}\frac{d^3\vec{p}}{(2\pi)^3}\frac{d^3\vec{q}}{(2\pi)^3}\frac{d^3\vec{w}}{(2\pi)^3}|f(\vec{k},\vec{p})|^2\times
\nonumber\\
&&\sum_{t,m}\sum_{s,r}\bar{v}^t(q)u^m(w)\bar{u}^s(k)v^r(p)\langle \vec{q},t;\vec{w},m|\vec{k},s;\vec{p},r\rangle.
\label{res546645}
\end{eqnarray}
Then,
\begin{eqnarray}
\langle \vec{q},t;\vec{w},m|\vec{k},s;\vec{p},r\rangle=2E_p2E_k\delta_{t,r}\delta_{m,s}(2\pi)^3\delta(\vec{k}-\vec{w})(2\pi)^3\delta(\vec{p}-\vec{q}),
\label{form66577}
\end{eqnarray}
such that Eq. (\ref{res546645}) becomes:
\begin{eqnarray}
\langle \Phi_0|\Phi_0\rangle=\int \frac{d^3\vec{k}}{(2\pi)^3}\frac{d^3\vec{p}}{(2\pi)^3}|f(\vec{k},\vec{p})|^2\times\sum_{r,s}\bar{v}^r(p)u^s(k)\bar{u}^s(k)v^r(p).
\label{res221332}
\end{eqnarray}
Using (we use the conventions in \cite{Srednicki} with $g^{\mu\nu}=(-1,1,1,1)$):
\begin{eqnarray}
&&\sum_su^s(k)\bar{u}^s(k)=-\gamma^{\mu}k_{\mu}+m
\nonumber\\
&&\sum_r\bar{v}^r(p)\gamma^{\mu}v^r(p)=2p^{\mu}\times 4
\nonumber\\
&&\sum_r\bar{v}^r(p)v^r(p)=(-2m)\times 4
\label{usef554664}
\end{eqnarray}
we obtain:
\begin{eqnarray}
\langle \Phi_0|\Phi_0\rangle=\int \frac{d^3\vec{k}}{(2\pi)^3}\frac{d^3\vec{p}}{(2\pi)^3}|f(\vec{k},\vec{p})|^2(4)(-1)[2p^{\mu}k_{\mu}+2m^2].
\label{finalres553663}
\end{eqnarray}

Similarly one finds:
\begin{eqnarray}
&&\langle \Phi_1|\Phi_1\rangle=\int \frac{d^3\vec{k}}{(2\pi)^3}\frac{d^3\vec{p}}{(2\pi)^3}|f(\vec{k},\vec{p})|^2\times
(-1)\sum_{r,s}\bar{v}^r(p)\gamma^5u^s(k)\bar{u}^s\gamma^5(k)v^r(p)=
\nonumber\\
&&=\int \frac{d^3\vec{k}}{(2\pi)^3}\frac{d^3\vec{p}}{(2\pi)^3}|f(\vec{k},\vec{p})|^2(4)(-1)[2p^{\mu}k_{\mu}-2m^2].
\label{res221332}
\end{eqnarray}

The two expressions for the fields $\Phi_0$ and $\Phi_1$ are completely identical apart from the factors:
\begin{eqnarray}
&&A=[2p^{\mu}k_{\mu}+2m^2]
\nonumber\\
&&B=[2p^{\mu}k_{\mu}-2m^2]
\label{factors55343}
\end{eqnarray}
where $A$ corresponds to $\Phi_0$ and B to $\Phi_1$.

One can further write:
\begin{eqnarray}
&&A=2p^{\mu}k_{\mu}+2m^2=(p+k)^2-p^2-k^2+2m^2=(p+k)^2+4m^2
\nonumber\\
&&B=2p^{\mu}k_{\mu}-2m^2=(p+k)^2-p^2-k^2-2m^2=(p+k)^2,
\label{calc66453}
\end{eqnarray}
where we used the fact that for the spinor solutions of the equation of motion $p^2=-m^2$ and $k^2=-m^2$ where $m$ is the  mass of the quark.

Since generically one can consider $p+k$ the momentum of the scalar and pseudoscalar state (it is sufficient to consider the wave function for the composite field to determine that) one can make  a change of variable $p+k=q$, $k=q-p$ to find the Fourier modes of the scalar and pseudoscalar fields.

In order to have a meaningful theory with a Lagrangian and a Hamiltonian  we should be able to normalize simultaneously the states $\Phi_0$ and $\Phi_1$. Note that if $m=0$ the normalization of the states is the same and we are able to normalize them simultaneously. As $m$ becomes different than zero the vacuum condensate forms and the normalization of the states differs showing that the $\Phi_0$ and $\Phi_1$ have actually different masses.
Then a necessary condition for the theory to make sense is to have:
\vspace{1cm}

1)Case I
\begin{eqnarray}
&&A=(2p^{\mu}k_{\mu})_1+2m^2=x^2
\nonumber\\
&&B=(2p^{\mu}k_{\mu})_2-2m^2=x^2
\label{case235}
\end{eqnarray}
or,

2) Case II
\begin{eqnarray}
&&A=(2p^{\mu}k_{\mu})_1+2m^2=-x^2
\nonumber\\
&&B=(2p^{\mu}k_{\mu})_2-2m^2=-x^2.
\label{case233}
\end{eqnarray}
where we added subscript $1$ and $2$ to show that the overall energy and mass for the quadrivector $p+k$ is different for the two fields. Both conditions for each case must be satisfied simultaneously. Then for case I we pick $(2p^{\mu}k_{\mu})_1=x^2-2m^2$ and for case II we pick $(2p^{\mu}k_{\mu})_2=-x^2+2m^2$ to determine that in either of these situations:
\begin{eqnarray}
|2p^{\mu}k_{\mu}|=|2m^2-x^2|\leq 2m^2.
\label{res44356}
\end{eqnarray}
Then using the fact that:
\begin{eqnarray}
|2p^{\mu}k_{\mu}|=\Bigg|2\sqrt{m^2+\vec{p}^2}\sqrt{m^2+\vec{k}^2}-2\vec{p}\cdot\vec{k}\Bigg|\geq 2m^2
\label{res32678}
\end{eqnarray}
we determine that $|2p^{\mu}k_{\mu}|=2m^2$ and consequently $x^2=0$ or $x^2=4m^2$. We can choose both possibilities but the second one would lead to zero mass for the scalar and non zero for the pseudoscalar. We thus pick the first choice as corresponding to the reality. However in the case where one considers two quarks one with the constituent mass $m$ and the other with the constituent mass $-m$ the second choice is the valid one and leads to a zero mass for the pseudoscalar. The second choice is also equivalent to considering both quarks with the same imaginary mass $im$.  We shall discuss the second choice from a different point of view in more details in section IV.

Then from  either of the Eq. (\ref{case235}) and (\ref{case233}) we determine:
\begin{eqnarray}
&&A=(2p^{\mu}k_{\mu})_1+2m^2=0=(p+k)^2+4m^2.
\nonumber\\
&&B=(2p^{\mu}k_{\mu})_2-2m^2=0=(p+k)^2.
\label{case23}
\end{eqnarray}
Since $(p+k)$ represents the four momentum of the scalar and pseudoscalar states we obtain that after spontaneous symmetry breaking the mass of the scalar $\Phi_0$ will become $4m^2$ whereas the pseudoscalar will be massless. This is quite a generic result which depends little or at all on the exact structure of the scalars and pseudoscalars in terms of the constituent quarks. Note that the factors of zero in the wave packet should be compensated by additional factors to be introduced in the denominators, more exactly in the structure function $f(\vec{k},\vec{p})$. Then the wave function for the composite scalars and pseudoscalars is constructed regularly afterwards with the new energy factors just determined through this method.

In summary we start with a picture where the fields $\Phi_0$ and $\Phi_1$ have the same mass be that zero to find out that the apparition of a quark condensate and the normalization of states leads to a change of picture where the scalar gains a different mass and the pseudoscalar becomes massless. In consequence we gave another proof of the Goldstone theorem applicable to composite theories.

\section{A different perspective on the mass eigenstates}

Assume instead of considering the initial Lagrangian in terms of the fields $\Phi_0$ and $\Phi_1$ we consider it in terms of the eigenstates (of the initial Lagrangian) $S_1\propto u_R^{\dagger}u_L$ and $S_2=S_1^{\dagger}\propto u_L^{\dagger}u_R$. Initially we have:
\begin{eqnarray}
&&\langle S_1^{\dagger}|{\cal H}|S_1\rangle=\vec{p^2}+m^2
\nonumber\\
&&\langle S_2^{\dagger}|{\cal H}|S_2\rangle=\vec{p^2}+m^2
\label{res665890}
\end{eqnarray}
with all the other matrix elements zero (here we took into account the normalization of states in order to get energy squared on the right hand side). Thus the hamiltonian is diagonal on the states $S_1$ and $S_2$. Assume that the quarks get a mass different than zero. Then a vacuum condensate forms because the element,
\begin{eqnarray}
\sum_{s,r}\bar{v}^r(q)(\frac{1-\gamma^5}{2})u^s(k)\bar{u}^s(k)(\frac{1-\gamma^5}{2})v^r(q)\approx (-8)m^2
\label{r665487}
\end{eqnarray}
is different than zero. Then the matrix elements in Eq. (\ref{res665890}) remain the same and  new matrix elements different than zero emerge: $\langle S_2^{\dagger}|{\cal H}|S_1\rangle=\langle S_1^{\dagger}|{\cal H}|S_2\rangle=a^2$. We do not know exactly the value of $a$ but we can diagonalize the hamiltonian to find that there are two eigenvalues:
\begin{eqnarray}
&&\lambda_1=\vec{p}^2+m^2-a^2
\nonumber\\
&&\lambda_2=\vec{p}^2+m^2+a^2.
\label{res546788}
\end{eqnarray}
We ask for the first eigenvalue to have the mass zero according to the Goldstone theorem. We obtain $a^2=m^2$ and $\lambda_2=p^2=\vec{p}^2+2m^2$. The eigenstates in this case are exactly $\Phi_1=\frac{S_1-S_2}{\sqrt{2}}$ with the mass zero and $\Phi_2=\frac{S_1+S_2}{\sqrt{2}}$ with the mass $2m^2$. Note that this is exactly the result that one would have obtained from a linear sigma model with a wrong sign mass term and a $\Phi^4$ interaction. In this case however we did not depart from the purely kinetic plus mass term.

\section{The scalar and pseudoscalar wave functions}

Let us consider again the $SU(2)_L \times SU(2)_R$ model mentioned in section II after the dynamical breaking of the symmetry through the formation of a vacuum condensate.  For illustration we just pick two states $\pi^+\propto \bar{d}\gamma^5u$ and $a^+\propto\bar{d}u$.  We know that after symmetry breaking the pion should be massless so the question that arises is:  what kind of quark substructure should $\pi^+$ and $a^+$ have to lead naturally to the desired mass spectrum? An example can be the following:
\begin{eqnarray}
&&\pi^+\approx \int d^4 y \bar{d}(x+y)\gamma^5u(x-y)
\nonumber\\
&&a^+\approx \int d^4 y \bar{d}(x+y)u(x-y)
\label{states4342}
\end{eqnarray}
Note that we shall consider one  quark with constituent mass $m$, the other with constituent mass $-m$ to account for usual mechanism of spontaneous symmetry breaking where the particles have the wrong mass term in the Lagrangian.
We consider that both quark states satisfy the Dirac equation with the mentioned masses. Then we apply the operator $\partial^{\mu}\partial_{\mu}$ to the pion (here we use $g^{\mu\nu}=(1,-1,-1,-1))$:
\begin{eqnarray}
&&\partial^{\mu}\partial_{\mu}\pi^+\approx
\nonumber\\
&&\approx \int d^4 y [(\partial^{\mu}\partial_{\mu})\bar{d}(x+y)\gamma^5u(x-y)+ \bar{d}(x+y)\gamma^5\partial^{\mu}\partial_{\mu}u(x-y)]+
\nonumber\\
&&\int d^4 y [2(\partial^{\mu}\bar{d})(x+y)\gamma^5\partial_{\mu}u(x-y)]=
\nonumber\\
&&-2m^2\int d^4 y \bar{d}(x+y)\gamma^5u(x-y)-\int d^4 y [2(\partial^{\mu}\bar{d})(x+y)\gamma^5\partial_{\mu}u(x-y)].
\label{res9973546}
\end{eqnarray}
We will treat the last term on the right hand side of Eq. (\ref{res9973546}) separately.  We use:
\begin{eqnarray}
2g^{\mu\nu}=2\gamma^{\mu}\gamma^{\nu}-[\gamma^{\mu},\gamma^{\nu}]
\label{g54664}
\end{eqnarray}
to write,
\begin{eqnarray}
&&\int d^4 y [2(\partial^{\mu}\bar{d})(x+y)\gamma^5\partial_{\mu}u(x-y)]=
\nonumber\\
&&\int d^4 y [2g^{\mu\nu}(\partial_{\mu}\bar{d})(x+y)\gamma^5\partial_{\nu}u(x-y)]=
\nonumber\\
&&\int d^4 y [2(\partial_{\mu}\bar{d})(x+y)\gamma^5\gamma^{\mu}\gamma^{\nu}\partial_{\nu}u(x-y)]-
\nonumber\\
&&\int d^4 y [(\partial_{\mu}\bar{d})(x+y)\gamma^5[\gamma^{\mu},\gamma^{\nu}]\partial_{\nu}u(x-y)]=
\nonumber\\
&&+2m^2\int d^4 y \bar{d}(x+y)\gamma^5u(x-y)-\int d^4 y [(\partial_{\mu}\bar{d})(x+y)\gamma^5[\gamma^{\mu},\gamma^{\nu}]\partial_{\nu}u(x-y)].
\label{res539076}
\end{eqnarray}
Here the minus sign in the first term on the right hand side of the equation comes from the anticommutator of $\gamma^{\mu}$ with $\gamma^5$.
The last term on the right hand side of Eq, (\ref{res539076}) is:
\begin{eqnarray}
&&\int d^4 y [(\partial_{\mu}\bar{d})(x+y)\gamma^5[\gamma^{\mu},\gamma^{\nu}]\partial_{\nu}u(x-y)]=
\nonumber\\
&&\int d^4 y \int \frac{d^4p}{(2\pi)^4} \frac{d^4k}{(2\pi)^4}\exp[ip(x+y)]\exp[ik(x-y)]\bar{d}(p)\gamma^5[\gamma^{\mu},\gamma^{\nu}]u(k)p_{\mu}k_{\nu}=
\nonumber\\
&&\int d^4 y \int \frac{d^4p}{(2\pi)^4} \frac{d^4k}{(2\pi)^4}\delta(p+k)(-p_{\mu}p_{\nu})\bar{d}(p)\gamma^5[\gamma^{\mu},\gamma^{\nu}]u(k)=0,
\label{proof657}
\end{eqnarray}
so it cancels.

By adding the results in Eqs. (\ref{res9973546}) and (\ref{res539076}) we obtain:
\begin{eqnarray}
\partial^{\mu}\partial{\mu}\pi^+=0
\label{res4343}
\end{eqnarray}
showing that the pseudoscalar is massless. Similarly one can show for $a^+$:
\begin{eqnarray}
\partial^{\mu}\partial_{\mu}a^+=-4m^2a^+,
\label{res4343}
\end{eqnarray}
signaling that the mass of $a^+$ is twice the absolute mass of an individual quark.

\section{Conclusions}

In this paper we discussed  the formation of a vacuum condensate and spontaneous symmetry breaking from the point of view of the meson substructure in terms of the constituent quarks.  The main result of the paper is to give another proof of the Goldstone theorem for composite states based on the normalization of the wave packets associated to the  mesons. Although we considered a simple $U(1)_L\times U(1)_R$ meson model our arguments can be extended easily to more flavors and larger symmetries by simply  computing  matrix elements  corresponding  to the meson states. We were thus able to confirm once more the standard features of spontaneous symmetry breaking.  One important remark is in order. In our work we showed that spontaneous symmetry breaking occurs as soon as the quarks or mesons gain constituent masses. However the order in which this happens is not clear cut.    In our simple model we showed that it is impossible for massive mesons to survive at any scale without spontaneous symmetry breaking.  Consequently the picture in which  confinement and chiral symmetry breaking occur at the same scale might be the correct one.

We relied on the quantum properties of the meson states in the theory and on the corresponding symmetry and we did not use  the $\Phi^4$ term present in the linear sigma model Lagrangian. Of course in order to give an accurate description of the reality one would need to take into account also the explicit symmetry breaking terms.

We also discussed the mass eigenstates of the Hamiltonian after spontaneous symmetry breaking  and briefly described a possible meson wave function that leads to the correct spectrum of states.

\section*{Acknowledgments} \vskip -.5cm

The work of R. J. was supported by a grant of the Ministry of National Education, CNCS-UEFISCDI, project number PN-II-ID-PCE-2012-4-0078.

\end{document}